\documentclass[12pt]{book}

\usepackage[dvips]{graphicx,color}
\usepackage{makeidx,tsukuba}

\makeauthorindex
\makeindex

\begin{document}

\BookTitle{\itshape The 28th International Cosmic Ray Conference}
\CopyRight{\copyright 2003 by Universal Academy Press, Inc.}
\pagenumbering{arabic}

\chapter{Atmospheric and Galactic Production and Propagation of Light 
Antimatter Nuclei}

\author{ Baret B.$^1$, Duperray R.$^1$, Boudoul G.$^1$, Barrau A.$^1$, Derome L.$^1$, Maurin D.$^2$, 
Protasov K.$^1$, and Bu\'enerd M.$^1$ \\
{\it 
(1) LPSC, Avenue des Martyrs 53, F-38026 Grenoble-cedex, France \\
(2) SAp CEA-Saclay, F-91191 Gif-sur-Yvette CEDEX, France
} \\
}

\section*{Abstract}
{\small\noindent 
The production and propagation of light antimatter nuclei $\overline{p}$, $\overline{d}$, 
$\overline{t}$, $\overline{^3He}$, $\overline{^4He}$ has been calculated using inclusive 
$\overline{p}$ production cross sections from a new data analysis, and coalescence models
for the production of composite particles. Particles were propagated using recently proven 
phenomenological approaches. The atmospheric secondary flux is evaluated for the first time. 
The Galactic flux obtained are larger than those obtained previously in similar calculations. 
The non-annihilating scattering contributions of the propagated particles are introduced. The 
preliminary results are shown and discussed.}

\section{Introduction}
A good knowledge of the Galactic production of (secondary) antiprotons and light antimatter nuclei, 
is a mandatory step on the way to the use of these (primary) particles in the search for exotic 
astrophysical sources, like dark matter or primordial black holes [1]. These calculations all rely 
on: a) The empirical knowledge of the antiproton (antinucleon) production cross sections since no 
theory can predict them with the required accuracy to date; and b) On phenomenological models for 
the production of composite particles using these cross sections. In this work, the production of 
Galactic antimatter has been revisited on the basis of a new analysis of the experimental inclusive 
$\overline{p}$ production cross sections available, and of a reliable approach of the particle 
propagation with parameters tightly constrained by Cosmic Ray data.

The atmospheric production of secondaries is a source of background for all embarked experiments 
(as well as a source of particles of high interest for other scientific fields). The 
knowledge of the corresponding flux in the Earth environment, is also a requirement for the 
interpretation of the satellite data which will be measured by the new generation of experiments.  

\section{Inclusive $\overline{p}$ production cross sections}
A selection of 654 data points of differential cross sections available from various 
$pA\rightarrow pX$ experiments over a broad range of incident energies (typically 10-450 GeV), has 
been analyzed consistently and fit using an analytical formula accounting for the cross section 
dependence on the reduced energy $x^*=E^*/E_{max}^*$, the transverse momentum (p$_t$), and the nuclear 
mass of the target [2].

\section{Coalescence models and antimatter nuclei}
The light antinuclei production cross sections were obtained by means of the microscopic coalescence 
model described in [3]. The simple coalescence model was also used for comparison. 

Recent nuclear collision data have accumulated experimental informations on the light nuclei
and antinuclei production cross sections for various systems over a broad energy range. Some features
particularly relevant to the present study are the quasi incident energy independence of the 
coalescence parameter for light nuclei production in $p+A$ collisions, and the observed equality of 
matter and antimatter coalescence parameters [4].
A sample of available $\overline{d}$ data have been analyzed in this context, using the standard 
coalescence model. The results support the conclusions of the quoted reference. They will be reported 
later [8].

\section{Particle production and propagation}

\subsection{In the Earth atmosphere}
\begin{figure}[hbt]
  \vspace{-0.2cm}
  \begin{center}
  \vspace{-1cm}
    \includegraphics[width=7cm]{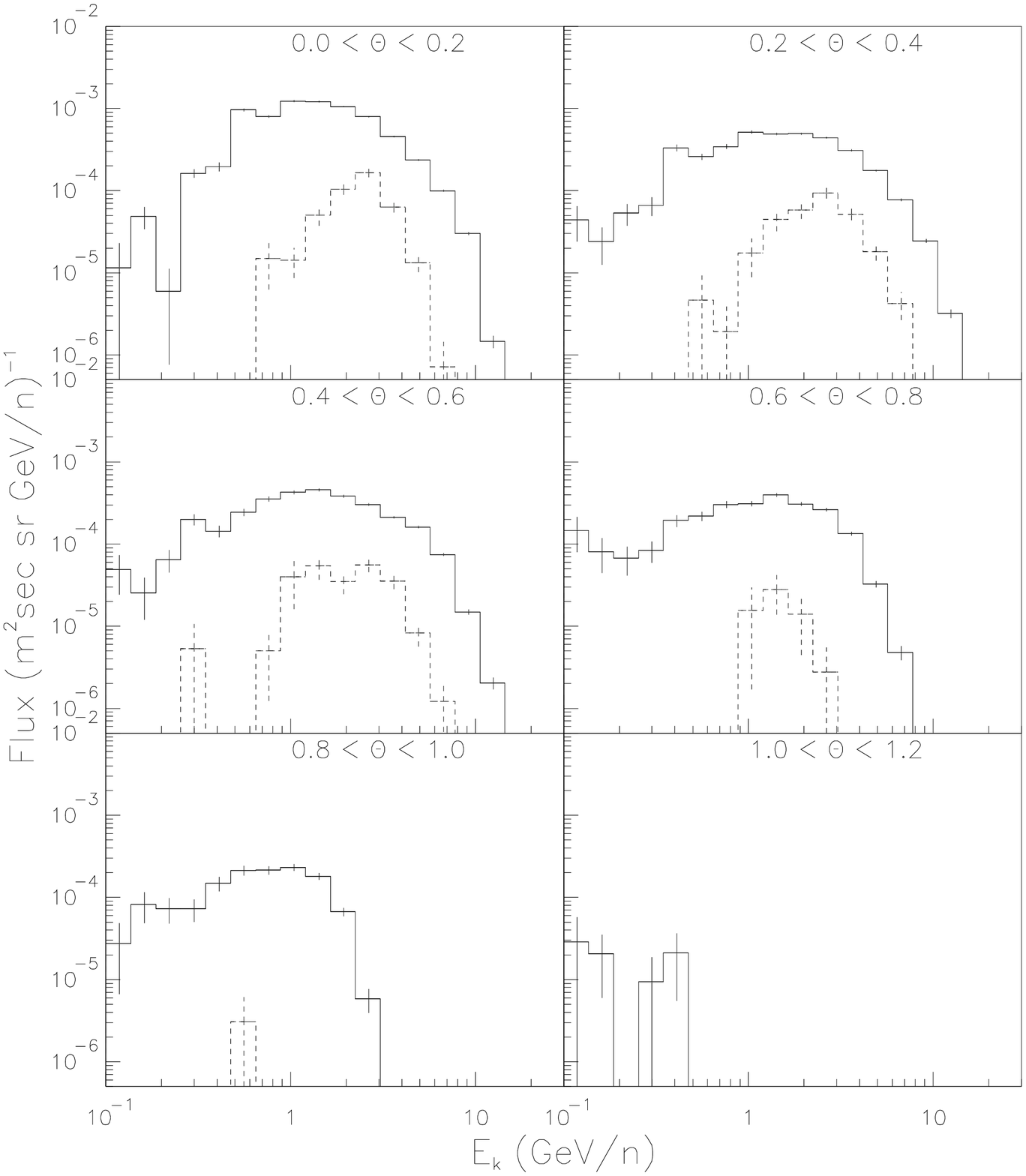} \includegraphics[width=7cm]{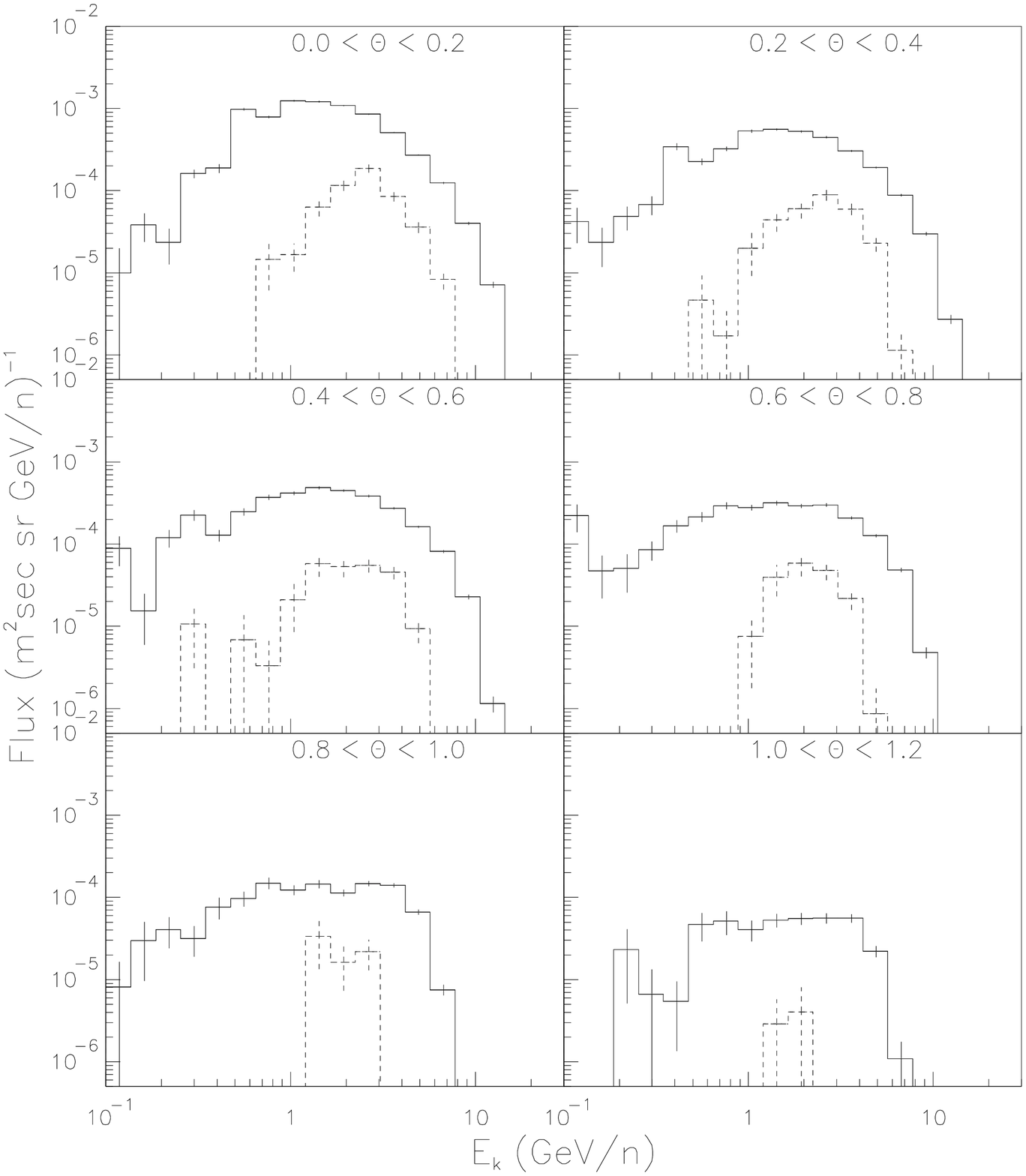}
  \end{center}
  \vspace{-0.5cm}
  \caption{\small\sl 
Simulation results for the atmospheric $\overline{p}$ (solid line) and $\overline{d}$ 
(dashed line) flux for the bins of latitude given on the figure. The $\overline{d}$ flux are 
multiplied by 2 10$^4$ for convenience.}
\end{figure}
The secondary antinuclei production in the Earth atmosphere was generated and propagated using a 
computer program developed recently and widely proven on various data samples (see [5] and refs 
therein). The resulting flux could be calculated at various coordinate positions covering the usual 
range of balloon and satellite experiments. The preliminary results for the $\overline{p}$ and 
$\overline{d}$ flux are shown on Fig.~1 for several bins of latitude between the equator and polar 
regions, at the AMS altitude ($\sim$~380~km). The left and right panels correspond to the incoming 
and outgoing flux respectively.

The $\overline{p}$ momentum spectra are peaked around 2~GeV per nucleon, which roughly corresponds 
to the velocity of the nucleon-nucleon center of mass system in which particles are produced with a 
maximum cross section at zero momentum, the incident energy times $\overline{p}$ production cross 
section distribution being peaked around 15~GeV. This spectrum is somewhat transformed by the 
cascade and transport in the atmosphere and in the Earth magnetic field, however.  

\subsection{In the Galaxy}
\begin{figure}[hbt]
  \vspace{-1cm}
  \begin{center}
  \includegraphics[width=7cm]{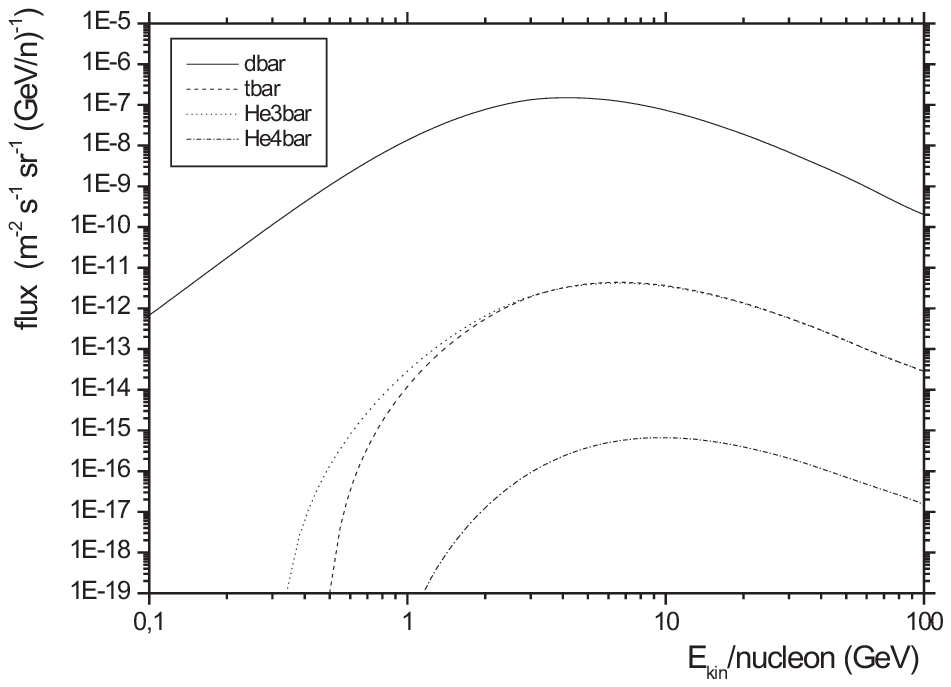} \includegraphics[width=7cm]{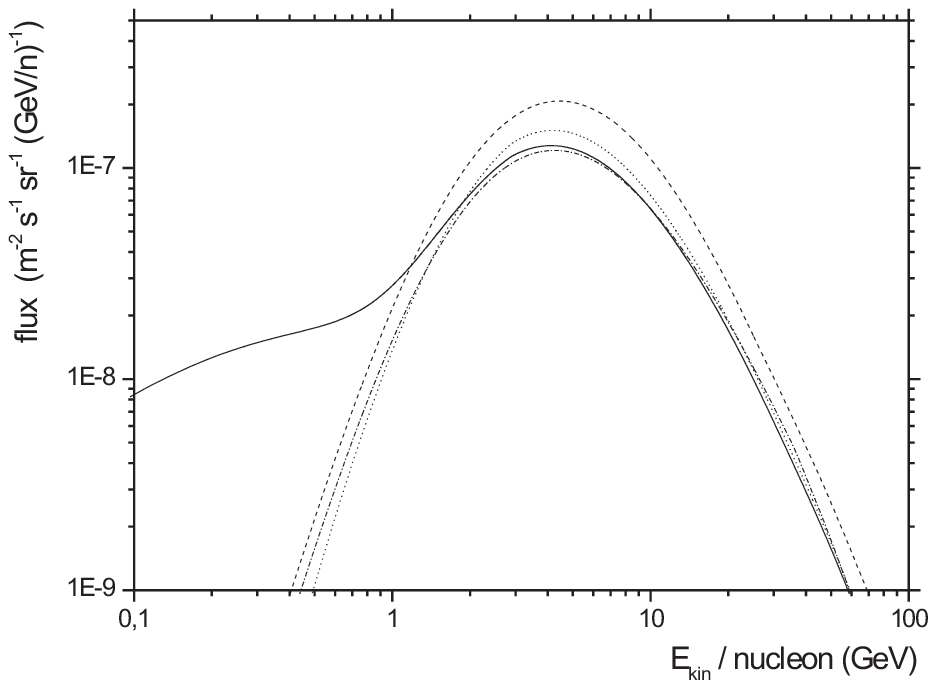}
  \end{center}
  \vspace{-0.5cm}
  \caption{\small\sl 
Left: Calculated Galactic flux for $\overline{d}$ (solid line), $\overline{t}$ (dashed 
line), $\overline{^3He}$ (dotted line), and $\overline{^4He}$ (dash-dotted line) using production 
cross sections from the standard coalescence model and the LBM for the propagation. Similar results 
are obtained with the microscopic coalescence model. Right: Same as left for $\overline{d}$ (solid 
line) and with contribution of non-annihilating rescattering of the particles (dashed line), showing 
the dramatic increase of the flux at low energy expected from this effect. The dot-dashed and dashed 
lines correspond to the minimum and maximum flux respectively, constrained by the nuclear CR data [6],
calculated in the Diffusion Model.}
\end{figure}
The Galactic propagation of particles was performed using both the standard Leaky Box Model (LBM), 
and the code developed by the LAPTH Annecy group, based on the Diffusion Model (DM) including
convection, reacceleration and energy losses, with the propagation parameters constrained by the B/C 
ratio [6]. Fig.~2 shows the preliminary results. The left panel shows the calculated Galactic 
flux for a few light antimatter nuclei, $\overline{d}$, $\overline{t}$ and $\overline{^3He}$, and 
$\overline{^4He}$. The flux were calculated at TOA using the LBM, with solar modulation effects 
included ($\Phi=500~MV$). 
In the search of primordial antimatter, $\overline{^4He}$ is the best candidate. It is thus of major 
importance to have reliable predictions for the Galactic secondary flux of this particle, to which 
future measurements could be confronted. 
Fig.~2 right shows the same LBM $\overline{d}$ flux with the effects of non-annihilating scattering of
the propagated particles included. This latter contribution was calculated by using a non-annihilating 
inelastic cross section based on an analytical form for the $pp\rightarrow pX$ differential cross 
section, renormalized using the total reaction cross section $\sigma_{dp}/\sigma_{pp}$ ratio. Note 
that this procedure is quite different than the usual ansatz. Details will be given in a forthcoming 
publication [8]. The DM calculations for $\overline{d}$ propagation are also shown on the figure 
(no rescattering effect included). They are in fair agreement with the LBM results, the latter being
between the limits calculated by the former. The non-annihilating inelastic contribution in the DM 
approach, are being calculated.
Note that the calculated $\overline{d}$ flux is found to be significantly larger, by a factor of 
about 2.5 than those obtained previously [7]. 

The ratio of the maxima of the calculated Galactic ($\sim$10$^{-8} (m^2.s.sr.GeV/n)^{-1}$) and 
atmospheric $\overline{d}$ flux (see Figs.~1 and 2) is about 10. The atmospheric background should thus 
not be a major source of experimental uncertainty since in addition, the knowledge of the particle 
momentum allows to discriminate atmospheric secondary particles from the Galactic ones.
\noindent
The $\overline{d}$ spectrum is found to peak around 4~GeV/n. This value is compatible with both a 
$\approx$16~GeV total energy production threshold, and a peaking of the $\overline{d}$ producing 
incident protons in the 20~GeV range (see above). This is at variance however with the atmospheric 
$\overline{d}$ spectrum obtained, which is centered at a lower energy (2-3~GeV). This feature is being 
investigated.

\noindent
Detailed calculations are in progress and will be reported later.

\vspace{\baselineskip}
\re 
1.\ Maurin D., Donato F., Taillet R., and Salati P. \ 2001, Astrophys.J. 536, 172; 
    Barrau A. et al., 2003, Astron. \& Astrophys. 398, 403.
\re
2.\ Duperray R., Huang C.Y., Protasov K., and Bu\'enerd M. \ 2003, astro-ph/0305274, May 15, 2003.  %
\re
3.\ Duperray R., Protasov K., and Bu\'enerd M., \ 2003, nucl-th/0301103,     .    
\re
4.\ Arsenescu R. et al. (NA52), \ Nucl. Phys. A661(1999)177c.   
\re
5.\ Yong Liu, L. Derome, and M. Bu\'enerd, Phys. Rev. D67, 073022.  
\re
6.\ Maurin D., Donato F., Taillet R., Salati. P., Astrophys.J. 555(2001)585; 
    Maurin D., Taillet R., and Donato F., Astron. \& Astrophys. 394 (2002)1039 
\re
7.\ Chardonnet P., Orloff J., and Salati P., Phys. Lett. B409(1997)313; 
     Donato F., Fornengo N., and Salati P., Phy. Rev. D62, 043003(2000).
\re
8.\ Duperray R. et al., in preparation for Phys. Rev. D.

\endofpaper
\end{document}